\documentclass[amsfonts, amssymb, amsmath, superscriptaddress, showpacs, reprint]{revtex4-1}
\usepackage{times}
\usepackage{bm}
\usepackage{hyperref}
\usepackage{graphicx}
\usepackage{epsfig} 
\usepackage{color}

\def\xx{\mathbf x}

\def\bb{\mathbf B}

\def\ccc{\color{black}} %% Remove color

\begin{document}

\title{Kinetic intermittency in magnetized plasma turbulence}
\author{Bogdan Teaca}
\email{bogdan.teaca@coventry.ac.uk}
\affiliation{Applied Mathematics Research Centre, Coventry University, Coventry CV1 5FB, United Kingdom}
\author{Alejandro \surname{Ba\~n\'on Navarro}}
%\email{banon@physics.ucla.edu}
\affiliation{Department of Physics and Astronomy, UCLA, 475 Portola Plaza, Los Angeles, CA 90095-1547, USA}
\author{Daniel Told}
%\email{dtold@physics.ucla.edu}
\affiliation{Department of Physics and Astronomy, UCLA, 475 Portola Plaza, Los Angeles, CA 90095-1547, USA}
\author{Frank Jenko}
%\email{jenko@physics.ucla.edu}
\affiliation{Department of Physics and Astronomy, UCLA, 475 Portola Plaza, Los Angeles, CA 90095-1547, USA}
\begin{abstract}
We employ {\ccc a gyrokinetic formalism, in an interval ranging from the end of the fluid scales to the electron gyroradius, to study intermittency at kinetic scales for magnetized plasma turbulence. We measure for the first time the intermittency of the distribution functions, accounting for velocity space structures and correlations generated by linear (Landau resonance) and nonlinear phase mixing.} Electron structures are found to be strongly intermittent and dominated by linear phase mixing, while nonlinear phase mixing dominates the weakly intermittent ions. This is the first time spatial intermittency and linear phase mixing are shown to be self-consistently linked for the electrons and, as the magnetic field follows the intermittency of the electrons at small scales, explain why magnetic islands are places dominated by Landau damping in steady state turbulence. 
\end{abstract}
% Plasma turbulence; Gyrokinetics; Gyrofluid and gyrokinetic simulations; MHD waves, plasma waves, turbulence
%
%
\maketitle

%%%%%%%%%%%%%%%%%%%%%%%%%%%%%%%%%%%%%%%%%%%%%%%%%%%%%%%%%%%%
%%%%%%%%%%%%%%%%%%%%%%%%%%%%%%%%%%%%%%%%%%%%%%%%%%%%%%%%%%%%
{\em Introduction.---} 
Turbulence in astrophysical plasma occurs over a wide range of spatial scales ($\ell$) \cite{Bruno:2013p1931, Kiyani2015}. While large scale dynamics can be captured by fluid approximations \cite{Zhou:2004p21}, the physics of turbulence on scales comparable to the proton gyroradius and smaller require a kinetic description \cite{Marsch:2006p1930, Schekochihin:2009p1131, Howes:2008p1132, Howes:2011p1370}. The departure of turbulence from self-similarity (i.e. scale invariance of the dynamics) leads to intermittency, which impacts simultaneously the locality of energetic interactions between scales and the formation of structures in real space \cite{Frisch}. 

The energy exchanges between scales (i.e. the energy cascade) and {\ccc the phenomenon of} intermittency have the same underlying cause: the existence of nonlinear interactions. The nonlinear mixing can lead to phase correlations between fluctuations, which generates intermittency. In Fourier space, which provides a natural projection of the turbulent dynamics on a hierarchy of scales ($k\! \sim\! 1/\ell$), these correlations are contained {\ccc in} the complex phases of the Fourier modes. The way the phases are correlated with each other decides how {\ccc the energy of a mode is distributed in real space}, either in a statistically uniform manner or in a few localized patches. A random or any uncorrelated distribution of phases will show no spatial intermittency regardless of the shape of the spectra. 

{\ccc For kinetic turbulence and its rigorous gyrokinetic (GK)~\cite{Brizard:2007p11, Krommes:2012p1373} limit in strongly magnetized plasmas, the distribution functions of the plasma species represent the dynamical quantities of interest. The role of the self-consistent electromagnetic fields, obtained from moments of the particle's distributions, is to mediate the nonlinear interactions between structures in the distribution functions \cite{Tatsuno:2009p1096}. This represents the underlying mechanism for the development of intermittency at the kinetic level. The kinetic dynamics occur in a position-velocity phase space~\cite{Schekochihin:2008p1034} involving couplings between velocity space structures \cite{Hammett:1992p1538, Plunk:2011p1357,Plunk:2010p1360} in addition to those between spatial scales \cite{BanonNavarro:2011p1274}. Unlike in fluid representations, there is no mechanism at the kinetic level for the direct correlation of the phases of the fields and the intermittency exhibited by them is inherited from the dynamics of the distribution functions. }

{\ccc Turbulence is typically associated with the development of localized structures, such as eddies, filaments and sheets, with the energy at small scales being contained in a few energetic structures \cite{Zhdankin:2015p1842}. Intermittency can thus be seen as the tendency of small scales to be less volume filling than larger scales and we will show this aspect in relation to the magnetic field. At kinetic scales, the dissipation of electromagnetic fluctuations is found to occur in highly localized current sheets by a series of numerical studies \cite{Wan:2012p1837, Wan:2015p1754, Karimabadi:2013p1855, Tenbarge:2013p1730} and satellite observations of the solar wind \cite{Perri:2012p1840, Perri:2012p1844, Osman:2012p1559, Osman:2012p1825, Osman:2014p1841, Sundkvist:2007p1734}. At the same time, particle-wave resonance is known to exchange energy between the electromagnetic fields and the particle distributions \cite{Klein:2016p1922}, affecting the generation of phase space structures. Analyzing the intermittency and the velocity structures of the distribution functions is crucial for assessing the correct route to collisional dissipation, key to the solar wind heating problem~\cite{Bruno:2013p1931}.

{\ccc In this letter, we study intermittency at kinetic scales. For the first time, we compute structure functions directly on the ion and electron distributions, which account for velocity space fluctuations and their mixing.} We show that magnetic intermittency follows that of the electrons and we link the emergence of intermittent electron structures with linear phase mixing (Landau resonance), a purely kinetic effect.  }

%%%%%%%%%%
\begin{figure*}[t]
\centering
\includegraphics[width = 0.95\textwidth]{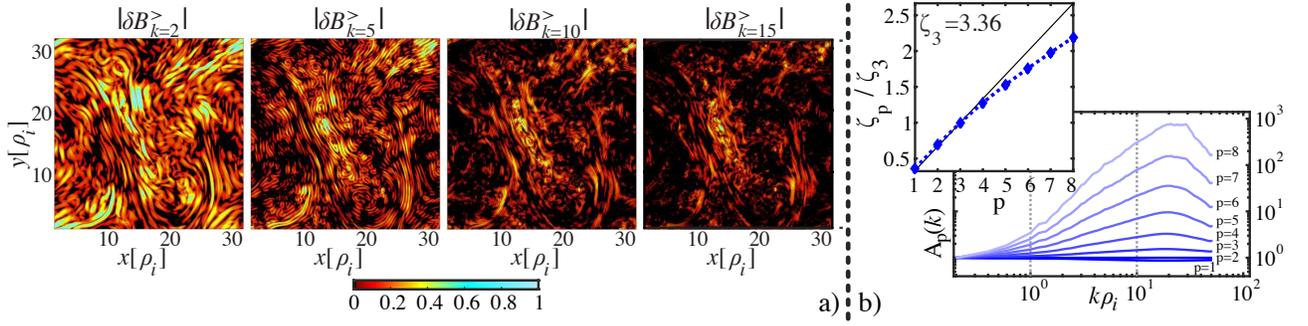} 
\caption{(color online) a) Real space visualization of the norm of perpendicular magnetic fluctuations. In all four panels, the same slice through the $z$ direction is taken, the magnetic field is normalized to its respective maximal value and values less than $0.1\%$ of the maximum are set to zero (black color). The four panels show the real space data, high-pass filtered beforehand in Fourier space (here $k_{\max}=51$ in units of $1/\rho_i$). The fact that small scales are less space filling is evident by the progressive increase in the black color. b) $A_p(k)$ for the perpendicular magnetic field and the $\zeta_p$ determined in the interval delimited by the two dotted lines by linear regression.}
\label{fig_Bperp}
\end{figure*}
%%%%%%%%%%

%%%%%%%%%%%%%%%%%%%%%%%%%%%%%%%%%%%%%%%%%%%%%%%%%%%%%%%%%%%%
%%%%%%%%%%%%%%%%%%%%%%%%%%%%%%%%%%%%%%%%%%%%%%%%%%%%%%%%%%%%
{\em Simulation details.---} 
In this study we use gyrokinetic simulations of magnetized proton-electron plasmas. This formalism assumes low frequencies (compared to the ion, here proton, cyclotron frequency) and small fluctuation levels {\ccc to remove the particle's fast gyro-motion, effectively reducing the relevant phase space to five-dimensions \cite{Brizard:2007p11, Howes:2006p1280}. While it neglects cyclotron resonance, gyrokinetics captures \cite{Told:2016p1915} the crucial dynamics of  kinetic Alfv\'en wave (KAW) turbulence in three spatial dimensions~\cite{podestasp13}.} The nonlinear gyrokinetic system of equations is solved with the Eulerian code {\sc GENE}~\cite{gene}. The data used in this Letter is taken from the simulation presented in Ref.~\cite{Told:2015p1712}, and it is briefly summarized in the following: The physical parameters of the simulations are chosen to be close to the solar wind conditions at 1 AU, with $\beta_{i}=8\pi n_{i}T_{i}/B_{0}^{2}=1$ and $T_{i}/T_{e}=1$. Proton and electron species are included with their real mass ratio of $m_{i}/m_{e}=1836$. The electron collisionality is chosen to be $\nu_{e}=0.06\, \omega_{A0}$ (with $\nu_{i}=\sqrt{m_{e}/m_{i}}\nu_{e}$), and $\omega_{A0}$ being the frequency of the slowest Alfv\'en wave in the system.  The evolution of the gyrocenter distribution is tracked on a grid with the resolution $\{N_{x}, N_{y}, N_{z}, N_{v_{\parallel}}, N_{\mu},N_{\sigma}\}=\{768, 768, 96, 48, 15, 2\}$, where ($N_x,N_y$) are the perpendicular, $(N_z)$ parallel, $(N_{v_{\parallel}})$ parallel velocity, and $(N_{\mu})$ magnetic moment ($\mu=mv^2_\perp/2B_0$) grid points, respectively. This covers a perpendicular {\ccc dealiased} wavenumber range of $0.2\le k_{\perp}\rho_{i}\leq 51.2$ (or $0.0047\leq k_{\perp}\rho_{e}\leq1.19)$ {\ccc in a domain $L_x=L_y=10\pi\rho_i$. In the parallel direction, a $L_z=2\pi L_{\parallel}$ domain is used, where $L_{\parallel} \gg \rho_i$ is assumed by the construction of gyrokinetic theory. A velocity domain up to three thermal velocity units ($v^{th}_\sigma=\sqrt{2T_\sigma/m_\sigma}$) is taken in each direction.} Here, $\rho_\sigma= \sqrt{T_{\sigma} m_{\sigma}} c / e B$ with the species index $\sigma$. A perturbed approach is employed, $F_{\sigma}$ being a constant Maxwellian background distribution with background density $n_{\sigma}$ and temperature $T_{\sigma}$, around which  fluctuations develop in the gyro-center distribution functions $f_{\sigma}=f_{\sigma}(x,y,z,v_{\parallel},\mu,t)$. {\ccc The fluctuations in the system are driven via a magnetic antenna potential, which is prescribed solely at the largest scale and evolved in time according to a Langevin equation~\cite{tenBargecpc14}. }

%%%%%%%%%%%%%%%%%%%%%%%%%%%%%%%%%%%%%%%%%%%%%%%%%%%%%%%%%%%%
%%%%%%%%%%%%%%%%%%%%%%%%%%%%%%%%%%%%%%%%%%%%%%%%%%%%%%%%%%%%
{\em Diagnostics \& magnetic intermittency.---}
{\ccc While the main purpose of this work is to analyze the intermittency of the distribution functions, we first introduce the diagnostics employed using the magnetic field fluctuations (denoted by $\bb$) to provide both an example of our approach and reference existing electromagnetic intermittency works~\cite{Wan:2012p1837, Wan:2015p1754, Karimabadi:2013p1855, Tenbarge:2013p1730, Perri:2012p1840, Perri:2012p1844, Osman:2012p1559, Osman:2012p1825, Osman:2014p1841, Sundkvist:2007p1734, Osman:2014p1647, Hnat:2005p1833, Kiyani:2007p1831,  Wan16, Coburn:2014p1850, Coburn:2015p27}}. To quantify spatial intermittency in the direction perpendicular to the magnetic guide field, we use $k$ high-pass filtered quantities \cite{Frisch}; e.g. the high-pass perturbed magnetic field is simply defined as $\delta \bb_k^>(\xx)=\int_{|{\bf q}|>k} \widehat \bb({\bf q},z) e^{i(q_x x+q_y y)}d{\bf q}$. The perpendicular structure functions of order $p$ are now defined as $S_p(k)= \left< |\delta B^>_{x,k}(\xx)|^p+ |\delta B^>_{y,k}(\xx)|^p \right> $, where the angle brackets refer to real space averages. From this definition we see that the structure functions are related to the $L^p$-space norms ($[S_p(k)]^{1/p}$) for the scale filtered quantities.

For turbulence within an ideal inertial range, the structure functions are expected to scale with $k$ as $S_p(k)=C_p k^{-\zeta_p}$, where $\zeta_p=pm+\gamma_p$ and the coefficients $\gamma_p$ measure the degree of intermittency. In the absence of intermittency, for which the self-similarity of the fields is exact, the anomalous $\gamma_p$ coefficients are zero and the scaling of the structure functions depends only on their order $p$ and the unique scaling factor $m=\zeta_3/3$. In general, the $C_p$ and $\zeta_p$ coefficients for different orders $p$ do not need to be related to each other, nor be universal. While structure functions can be linked to the energy spectra and the scale flux of energy, the normalized structure functions, defined as
\begin{align}
A_p(k)= \frac{ S_p(k)}{[S_2(k)]^{p/2}}\; \label{Ap}
\end{align}
are more useful for the study of intermittency. In the absence of intermittency $A_p(k)$ is independent of $k$. This can be easily seen for any self-similar scale transformation, e.g. $\delta \bb_{\lambda k}^>\!=\!\lambda^{-m} \bb_k^>$. However, $k$ independence for $A_p(k)$ is also observed for a real space Gaussian distribution, which leads to the incorrect measure of intermittency as a departure from the Gaussian distribution, rather than a departure from scale invariance \cite{Hnat:2003p1829}. For strongly intermittent $k$-intervals, the $A_p(k)$ curves exhibit an explosive separation for different orders $p$.

In FIG.~\ref{fig_Bperp}-a) we depict the tendency for small scale structures to be less volume filling than larger scales, exemplified for the magnetic field. This is the main effect associated with intermittency in real space. In FIG.~\ref{fig_Bperp}-b) we plot the corresponding $A_p(k)$ for the magnetic field and the scaling index $\zeta_p${\ccc , determined by a linear regression of $\log(S_p)=-\zeta_p \log(k)+\log(C_p)$ in the interval $k\rho_i\in[1, 10]$. The qualitative intermittency results found for the magnetic field are consistent with multi fractal intermittency ($\zeta_p$ has an increased deviation from the diagonal line for larger $p$), supporting the idea of kinetic Alfv\'en wave (KAW) cascade \cite{Howes:2011p1370} at scales smaller than $\rho_i$. }

%%%%%%%%%%%%%%%%%%%%%%%%%%%%%%%%%%%%%%%%%%%%%%%%%%%%%%%%%%%%
%%%%%%%%%%%%%%%%%%%%%%%%%%%%%%%%%%%%%%%%%%%%%%%%%%%%%%%%%%%%
{\em On phase mixing dynamics.---} We concentrate hereon on the underlying kinetic dynamics. For GK turbulence, the five-dimensional dynamics involve the generation of small scales in the parallel (linear phase mixing \cite{Watanabe:2006p1444, Zocco:2011p1306, PLA:10293237}) and perpendicular (nonlinear phase mixing \cite{Tatsuno:2009p1096, Schekochihin:2008p1034, Hammett:1992p1538, Plunk:2011p1357, Plunk:2010p1360}) velocity directions. The nonlinear phase mixing occurs as part of the same nonlinear interactions that are responsible for the generation of small spatial structures and the emergence of intermittency, namely the advection by drift velocities of the nonadiabatic part of the gyro-center distribution function ($h_{\sigma}\!=\!f_{\sigma}+\![q_\sigma\bar \phi_{\sigma}\!+\!\mu \bar B_{\parallel \sigma}]F_{\sigma}/T_{\sigma}$; with $\phi$ the first order self-consistent electrostatic potential, $B_{\parallel}$ the first order magnetic fluctuation in the parallel direction and the overbar refers to a species dependent gyro-average~\cite{Merz2009}). Intermittency emerges solely as a result of these nonlinear interactions. The linear phase mixing term ($\sim\!\! \partial h_\sigma/\partial v_\parallel$) cannot generate intermittency, however, it is the term that leads to Landau Damping \cite{Plunk:2013p1890, Villani:2014p1545} and the generation of ever-smaller parallel velocity structures \cite{Chust:2009p1582, Bratanov:2013p1871, ChuLi:2015p1853}.

%%%%%%%%%%%%%%%%%%%%%%%%%%%%%%%%%%%%%%%%%%%%%%%%%%%%%%%%%%%%
%%%%%%%%%%%%%%%%%%%%%%%%%%%%%%%%%%%%%%%%%%%%%%%%%%%%%%%%%%%%
{\em Parallel velocity decomposition.---} 
In addition to spatial intermittency, we want to account for structures developed in the parallel velocity for the ion and electron non-adiabatic distribution functions, i.e. $h(v_\parallel)$, where we suppress dependencies to simplify the notations. We use a Hermite representation \cite{Hatch:2013p1869, Hatch:2014p1639, Loureiro:2013p1860, Numata:2015p1880, PLA:10293237}, employing the Hermite functions $\psi_n(v_\parallel)=  \left (2^n n! \sqrt{\pi} \right)^{-{1/2}} e^{v_\parallel^2/2} (-d/dv_\parallel)^n e^{-v_\parallel^2}$, which are orthonormal $\int_{-\infty}^{\infty} \! \psi_n(v_\parallel)\psi_m(v_\parallel)\, \mathrm{d}v_\parallel \!\!=\! \delta_{nm}$, with $\delta_{nm}$ the Kronecker delta.

The $n$-filtered parallel velocity is defined here as $h_n(v_\parallel)= \hat h_n \psi_n(v_\parallel)$, where $\hat h_n = \int_{-\infty}^{+\infty} h(v_\parallel) \psi_n(v_\parallel) dv_\parallel$ are Hermite amplitudes related to velocity moments of the distribution function; $n=\{0,1,2\}$ relate respectively to fluctuations of the particle density, bulk velocity and particle kinetic energy and are seen as fluid like contributions \cite{Loureiro:2013p1860, Numata:2015p1880, PLA:10293237}. The original distribution function is simply obtained as a sum over all possible $n$-filtered contributions, i.e. $h(v_\parallel)=\sum_{n=0}^\infty  h_n (v_\parallel)$. While $h_0 (v_\parallel)$ has a simple Gaussian form in $v_\parallel$, for ever larger values of the integer $n$ we select ever smaller scales in $v_\parallel$. Velocity scales represented by $n\ge3$ can be seen as kinetic only contributions that are not captured by simple fluid closures (e.g. Grad-13 \cite{CPA:CPA3160020403}) and are deemed to have $n$-independent dynamics \cite{Parker:1995p136}. To observe the influence of these kinetic only velocity structures, we designate $h_{\ge3}(v_\parallel)$ as the sum over all $n\ge3$ contributions, i.e. $h_{\ge3}(v_\parallel)=\sum_{n=3}^\infty  h_n (v_\parallel)$.

%%%%%%%%%%%%%%%%%%%%%%%%%%%%%%%%%%%%%%%%%%%%%%%%%%%%%%%%%%%%
%%%%%%%%%%%%%%%%%%%%%%%%%%%%%%%%%%%%%%%%%%%%%%%%%%%%%%%%%%%%
{\em Kinetic structure functions.---} The high-pass filters for $h_\sigma$, omitting the $\sigma$ species index to simplify the notations, are 
\begin{align}
\delta h_k^>(x,y,z,v_{\parallel},\mu)=\int_{|{\bf q}|>k} \!\!\!\!\!\widehat h({\bf q},z,v_{\parallel},\mu) e^{i(q_x x+q_y y)}d{\bf q} \;.
\label{deltaH}
\end{align}
The structure functions of order $p$ for a given species $\sigma$ are defined as $S_p(k)= \int  |\delta h_k^>(x,y,z,v_{\parallel},\mu)|^p dV$, where $dV=dx\, dy\, dz\, dV_v$ is the five-dimensional phase space volume element with $dV_v={2\pi B_0}/{m_\sigma}dv_{\parallel}d\mu$ the velocity space volume element for the GK problem. Considering $|\delta h^>_k|$ ensures that velocity space cancellations do not occur during integration. Here, we take a single given plane in $z$. The definition of $A_p(k)$ is still given by Eq.~(\ref{Ap}), with the same interpretation. 

%%%%%%%%%%
\begin{figure}[t]
\centering
\includegraphics[width = 0.43\textwidth]{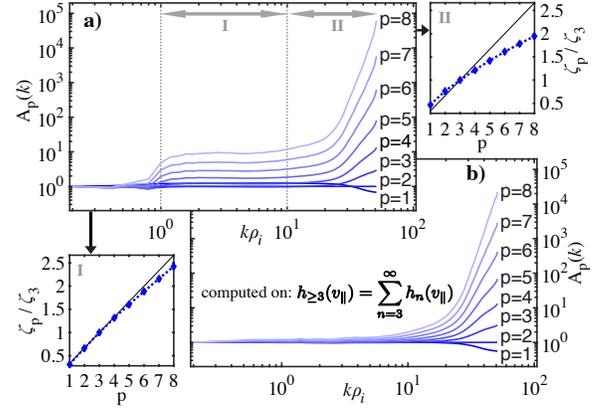} %44
\caption{a) Normalized structure functions of orders $p=1..8$, for the ion (proton) distribution function. The curves are normalized to their respective value at $k\rho_i=0.2$. The associated panels (I \& II) plot $\zeta_p$ determined by linear regression in the intervals indicated. b) The same main plot computed on $h_{\ge3}(v_\parallel)$.}
\label{fig_Api}
\end{figure}
%%%%%%%%%%

%%%%%%%%%%
\begin{figure}[b]
\centering
\includegraphics[width = 0.43\textwidth]{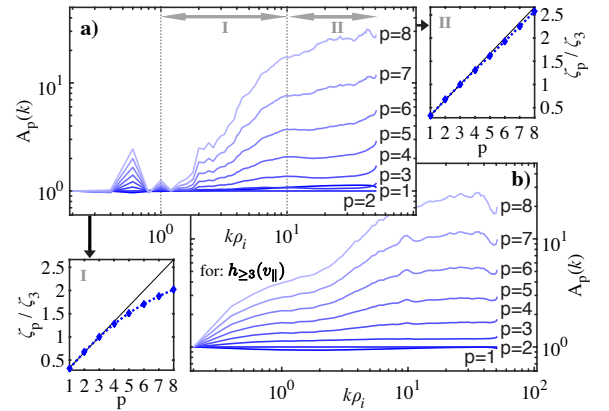}
\caption{Normalized structure functions for electron distribution function. Same remarks listed in  the caption of FIG.~\ref{fig_Api} apply.}
\label{fig_Ape}
\end{figure}
%%%%%%%%%%

%%%%%%%%%%
\begin{figure}[t]
\centering
\includegraphics[width = 0.48\textwidth]{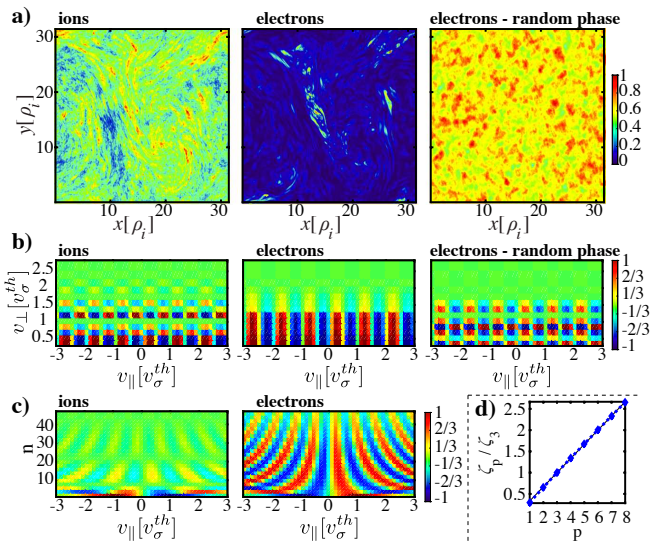}
\caption{a) Real space visualization in a $(x,y)$ plane at a given $z$, normalized to their respective in plane maximal value. b) velocity structures for $h_{21}$ at a given spatial point. c) the $\mu$ integrated $n$-scale structures as a function of $v_\parallel$. d) the $\zeta_p$ scaling for the interval I in FIG.~\ref{fig_Ape}, computed for the phase randomized electrons.}
\label{fig_s1}
\end{figure}
%%%%%%%%%%

We plot the normalized structure functions $A_p(k)$ computed for the ions (protons) in FIG.~\ref{fig_Api} and for the electrons in FIG.~\ref{fig_Ape}. For two wavenumber intervals, denoted by I \& II in the panel (a) of each figure, we show the $p$-scaling of the exponents $\zeta_p${\ccc, computed as for the magnetic field. Here, we are interested in the qualitative form of the $\zeta_p/\zeta_3$ curves and a refinement of the choice of intervals does not impact this aspect in a meaningful way}. In panels (b) for each figure, we show the same analysis performed on the kinetic only velocity scale contributions, given by $h_{\ge3}(v_\parallel)$, for which we mention that the corresponding $\zeta_p$ scalings remain qualitatively the same. 

For the fluid scales ($k \rho_{i}\!<\! 1$), we see an absence (or a strong attenuation) of intermittency. The ions show a strong intermittent behavior at small scales (captured by the divergence of $A_p(k)$ curves and the departure of the $\zeta_p$ scaling from the non-intermittency line). The electrons exhibit a multi-fractal intermittency in the range $k\rho_i\!\in\![1, 10]$ and a non-intermittent behavior at the smallest scales (captured best by the $\zeta_p$ scalings). These regimes seem to correspond with  the peak of free energy dissipation for each species (see FIG.~4 in Ref.~[\onlinecite{Told:2015p1712}]). 

At $k \rho_{i}\!\approx\! 1$ we see a break in the scaling of $A_p$ for the ions, which is absent in the analysis of the kinetic only velocity contributions $h_{\ge3}(v_\parallel)$. While this break does not influence the anomalous scaling of $\zeta_p$, it shows that the non-universal $C_p$ coefficients (e.g. $C_8\ne C_4^2$) are sensitive to fluid-like velocity contributions. The fluid-like velocity contributions have a stronger impact on the electrons, attenuating their intermittency at large spatial scales ($k \rho_{i}< 1$).

%%%%%%%%%%%%%%%%%%%%%%%%%%%%%%%%%%%%%%%%%%%%%%%%%%%%%%%%%%%%
%%%%%%%%%%%%%%%%%%%%%%%%%%%%%%%%%%%%%%%%%%%%%%%%%%%%%%%%%%%%
{\em Phase mixing and intermittency.---} 
Next, keeping only perpendicular spatial scales smaller than the ion gyroradius (i.e. $k \rho_i\! >\!1$), we look at velocity space structures developed by the ions and electrons. In addition, we consider the electron case for which in Eq.~(\ref{deltaH}) we randomize the Fourier mode phases, while keeping the same spectral energy density. By doing so we destroy the nonlinear correlations. In FIG.~\ref{fig_s1}-a) we plot the $\int  |h_{\ge3}|^2 dV_v$ for the three cases, visually observing that the electrons kinetic structures are highly intermittent compared to the ions and that they occupy a smaller volume. Randomizing the Fourier phases for the electrons destroys their real space structures and suppresses kinetic intermittency, as seen in FIG.~\ref{fig_s1}-d) for the scaling of $\zeta_p$ in the (previously multi-fractal) range $k\rho_i\!\in\![1, 10]$. 

For spatial points located in high intensity structures ($>~\!\!\!90\%$ value in the previous plots), we look in FIG.~\ref{fig_s1}-b) at the velocity space $(v_\parallel,v_\perp)$ for $h_{21}$. Selecting only one parallel velocity scale avoids cancellations or smudging from taking place during $v_\parallel$ integration and provides for a much clearer message. The $n\!=\!21$ parallel velocity scale determines the structure size in $v_\parallel$ and is here an arbitrary choice. The dominance of nonlinear phase mixing (the case for ions) can be seen as the structures in $v_\perp$ become apparent ($v_\perp\!=\!\sqrt{\mu}$ in normalized units). By comparison, the dominance of linear phase mixing for the electrons manifests as the absence of $v_\perp$ structures (except for an exponential decay for large values) and the emergence of long structures of given sign. We mention that non-intense points for the electrons exhibit velocity space structures similar to those seen for the ions, however, with an intensity smaller by one order of magnitude than points exhibiting linear phase mixing. The linear phase mixing role for the electrons is clear from FIG.~\ref{fig_s1}-c), as large scale velocity structures (low $n$) are smoothly transformed into smaller scales (here only odd values of $n$ are shown for clarity; the same picture would be observed if an initial intense large velocity scale structure would be tracked in time \cite{Klein:2016p1922}).

The most striking result is observed for the phase randomized electrons, for which the linear mixing in velocity space is destroyed. This supports the idea that linear phase mixing, which includes Landau Damping, occurs in intermittent structures. The nonlinear phase correlations that are responsible for the emergence of intermittency also ensure that Landau resonance is achieved, showing that a balance between the linear phase mixing and the nonlinear interactions emerges at sub-gyroradius scales, as proposed at large scales by Ref.~[\onlinecite{PLA:10293237}].

%%%%%%%%%%%%%%%%%%%%%%%%%%%%%%%%%%%%%%%%%%%%%%%%%%%%%%%%%%%%
%%%%%%%%%%%%%%%%%%%%%%%%%%%%%%%%%%%%%%%%%%%%%%%%%%%%%%%%%%%%
{\em Discussions and conclusions.---}
We analyzed large resolution GK simulations, pertinent to kinetic Alfv\'en wave turbulence, {\ccc for the first time analyzing the intermittency of the plasma distribution functions.} At the fundamental interaction level for turbulence, intermittency manifests itself as an increase in the nonlocal contributions made to the energetic interactions between two scales. The novel intermittency results presented here agree with the locality studies performed in the past for GK turbulence \cite{Teaca:2012p1415, Teaca:2014p1571, Told:2015p1712} and the apparent absence of unique locality exponents at particular scales, now identified as being strongly intermittent.

{\ccc For kinetic scale turbulence, we find nonintuitive results that highlight the importance of phase space dynamics}. The electrons are strongly intermittent at kinetic scales ($k \rho_i \ge1$) while the ions show little to no intermittent behavior in the same range. {\ccc The generation of intermittent structures for the electrons is caused by the same nonlinear term responsible for the cascade to small scales and the nonlinear phase mixing in velocity space. Yet, unlike the surrounding background, the most intense structures in the electron distribution function exhibit a clear parallel velocity structure, indicative of linear phase mixing being dominant. This shows evidence that nonlinear correlations play a role in obtaining a strong linear mixing resonance \cite{Klein:2016p1922}, which is achieved in intermittent structures for the electrons.} For the simulations analyzed here ($\beta=1$), the ion inertial scale corresponds to $\rho_i$ in value. {\ccc At scales smaller than the ion inertial scale, the magnetic field is decoupled from the ions and embedded in the electron flow, exhibiting structures similar to $h_e$  and developing a similar multi-fractal intermittent distribution of real space structures as depicted by the electrons.} As intense electron structures are dominated by linear phase mixing, this explains why magnetic islands are places dominated by Landau Damping \cite{Loureiro:2013p1860, Numata:2015p1880}.

%%%%%%%%%%%%%%%%%%%%%%%%%%%%%%%%%%%%%%%%%%%%%%%%%%%%%%%%%%%%
%%%%%%%%%%%%%%%%%%%%%%%%%%%%%%%%%%%%%%%%%%%%%%%%%%%%%%%%%%%%
\begin{acknowledgments}
{{\em Acknowledgments.---} We acknowledge the Max-Planck Princeton Center for Plasma Physics for facilitating the discussions that lead to this paper and would like to thank Paul Crandall and Chris Pringle for their comments on the manuscript. The research leading to these results has  received funding from the European Research Council under the European Unions Seventh Framework Programme (FP7/2007V2013)/ERC Grant Agreement No. 277870. The gyrokinetic simulations presented in this work used resources of the National Energy Research Scientific Computing Center, a DOE Office of Science User Facility supported by the Office of Science of the U.S. Department of Energy under Contract No. DE-AC02-05CH11231.}
\end{acknowledgments}

%%%%%%%%%%%%%%%%%%%%%%%%%%%%%%%%%%%%%%%%%%%%%%%%%%%%%%%%%%%%
%%%%%%%%%%%%%%%%%%%%%%%%%%%%%%%%%%%%%%%%%%%%%%%%%%%%%%%%%%%%
%\bibliography{Reference_GK_Locality}
%
%

%%%%%%%%%%%%%%%%%%%%%%%%%%%%%%%%%%%%%%%%%%%%%%%%%%%%%%%%%%%%
%%%%%%%%%%%%%%%%%%%%%%%%%%%%%%%%%%%%%%%%%%%%%%%%%%%%%%%%%%%%
\end{document}